\documentclass[pre,twocolumn,showpacs,floatfix]{revtex4}

\usepackage{graphicx}

\begin{document}

\title{Self-assembly of polyhedral shells: A molecular dynamics study}

\author{D. C. Rapaport}
\email{rapaport@mail.biu.ac.il}
\affiliation{Physics Department, Bar-Ilan University, Ramat-Gan 52900, Israel}

\date{July 5, 2004}

\begin{abstract}

The use of reduced models for investigating the self-assembly dynamics
underlying protein shell formation in spherical viruses is described. The
spontaneous self-assembly of these polyhedral, supramolecular structures, in
which icosahedral symmetry is a conspicuous feature, is a phenomenon whose
dynamics remain unexplored; studying the growth process by means of computer
simulation provides access to the mechanisms underlying assembly. In order to
capture the more universal aspects of self-assembly, namely the manner in which
component shapes influence structure and assembly pathway, in this exploratory
study low-resolution approximations are used to represent the basic protein
building blocks. Alternative approaches involving both irreversible and
reversible assembly are discussed, models based on both schemes are introduced,
and examples of the resulting behavior described.

\end{abstract}


\maketitle

\section{Introduction}

Self-assembly at the molecular scale is a ubiquitous process in the natural
world and predicted to play a significant role in advancing nanotechnology. The
formation of the protein shells, known as capsids, that provide the packaging
for spherical (or polyhedral) viruses is a particularly familiar instance of
natural self-assembly; the rational design of antiviral agents would benefit
from an improved understanding of how such shells assemble, a phenomenon at the
border between biology and chemistry. Direct visualization of evolving
molecular assemblies is inherently difficult, since intermediate states elude
experimental capture, and final states reveal little about their assembly
pathways; thus the simulation of suitably designed models ought to prove
helpful in exploring the processes and pathways involved.

Virus capsids have highly symmetric shapes, reflecting the fact that they are
assembled from multiple copies of either a single molecular component -- the
capsomer -- or a small number of distinct capsomers \cite{bak99}. Capsid
assembly, a process whose details are little understood \cite{fox94}, is
governed by different classes of interactions: There are interactions between
capsomer proteins and nucleic acids (in the form of DNA or RNA) that initiate
and regulate the assembly process, and subsequently stabilize the packaged
genetic material inside the capsid. More significantly from the perspective of
the present study, there are also interactions between the capsomer proteins
that stabilize the shell structure itself. What makes capsid self-assembly an
ideal candidate for simulation, despite this apparent biochemical complexity,
is the fact that it is able to occur reversibly {\em in vitro} \cite{pre93}
(with scaffold proteins participating in the growth but not affecting
stability), without the genetic material that is essential to the virus {\em in
vivo} (in other cases the nucleic acid does play a stabilizing role
\cite{spe95}). In addition, structurally intact empty shells occur {\em in
vitro} after removal of their contents, and viruses themselves form empty
capsids \cite{cas62}; background information of this kind simplifies the model
design considerably, since the system need only consist of a very small number
of well-characterized component types. The goal of the present work is to use
molecular dynamics (MD) simulation \cite{rap04} in modeling the capsid assembly
process, based on suitably simplified representations of the essential
molecular components.

Motivation for simplified descriptions, that avoid becoming embroiled in the
detailed physical and chemical characteristics of the capsomers, stems from the
prominence of icosahedral symmetry among spherical viruses, irrespective of
their biological origins. Nature has adopted this structural motif (the other
basic design is a helical tube) precisely because the high degree of symmetry
leads to a minimal set of construction rules \cite{cri56}. Since the task of
the genetic information carried by the viral nucleic acid is not only to
instruct the virus how to infect the host, but also to specify how it must
replicate itself, if less information can be devoted the latter mission, more
will be available for the pernicious primary task. At the same time, shapes
based on icosahedra come close to maximizing the volume to surface ratio,
another advantageous feature. Beyond their basic packaging role, capsomers also
have an important function in the virus life cycle \cite{joh96} that the simple
structural models do not attempt to address.

Self-assembly is related to crystallization, in the sense that both are
governed by the laws of thermodynamics, with the obvious difference that while
crystals can, in principle, grow without limit, viral growth is self-limiting.
The processes are driven by bond formation between assembling units, whether
atoms or protein complexes, with the goal of reaching the relevant minimal
free-energy state. Assembly of symmetric structures occurs both in biological
and nonbiological contexts; while crystal growth may not be of particular
biological relevance, the formation of icosahedral fullerene molecules is an
indication that some geometries are common to both. Analogous icosahedral
motifs are to be found in geodesic domes; these owe their detailed structure
both to the same minimalist construction rules, as well as to considerations of
optimal rigidity. Nature embodies a great deal of what is considered successful
engineering design, and indeed, many noteworthy engineering achievements borrow
from the geometric forms found in nature, even though size scales and materials
can differ greatly.

In order to avoid a substantial, if not overwhelming, degree of incorrect
assembly, it seems plausible that the construction of viral capsids demands a
generalized scheme independent of many of the molecular details of individual
viruses; nonspecific assembly pathways might well exist, a knowledge of which
could be important in finding ways to modify or inhibit the construction
process, with obvious therapeutic and technological benefits. Consequently,
initial exploration should focus on the basic `shape' of the constituent
capsomers, clearly an important factor in determining the assembly outcome.
Such low-resolution descriptions amount to little more than caricatures, but,
if general organizational principles do exist, they could well be sufficiently
robust to reveal themselves in models from which extraneous detail has been
removed. There is a longstanding tradition of working with reduced models that
has proved extremely successful with other inherently complex phenomena, such
as micelle growth and protein folding,

The use of MD for modeling capsid self-assembly was introduced in a previous
study \cite{rap99} involving polyhedra built from 60 triangular units. In the
present paper the work is extended in two directions, first by showing how more
relevant trapezoidal capsomer shapes can be used and larger shells constructed,
and second by demonstrating alternative ways of designing capsomer models.
Despite the simplifications, such simulations provide access to assembly
pathways, and are able to predict time-dependent partial structure populations
that can, in principle, be compared with experiment \cite{end02}. The following
sections describe alternative model designs (Section~\ref{sec:capdes}), the
nature of the bond interactions (Section~\ref{sec:bondint}), different assembly
scenarios (Section~\ref{sec:assem}) including the pathways and supplementary
interaction rules involved, computational techniques (Section~\ref{sec:tech}),
and results (Section~\ref{sec:res}) from extensive simulations. It should be
stressed that while the focus is primarily on the formation of capsid shells,
the method is not in any way restricted to virus structures, and is applicable
to supramolecular assembly in general; a brief demonstration
(Section~\ref{sec:other}) of this capability precedes the concluding section.

\section{\label{sec:capdes}Capsomer design}

\subsection{General considerations}

Individual capsomers are large protein complexes; owing to their size and
complexity, such structures must be represented in a highly reduced form, while
retaining sufficient detail to ensure meaningful behavior when studied by MD
simulation. Viewed from this perspective, capsomers have two principal but not
entirely independent characteristics. One is the effective molecular shape;
this must be tailored to ensure capsomers fit together to form closed
polyhedral shells. The other concerns the interactions between adjacent
capsomer regions in the final structure; these are responsible for driving
self-assembly and maintaining the structural integrity of the finished shell,
and must be defined accordingly.

There are few guidelines to aid in the model design. While general
thermodynamic considerations can help choose ranges for the force parameters,
the approach itself is entirely empirical. Progress in this kind of
discrete-particle modeling is tied to the advance in available computer power,
with increasing computational capability allowing the incorporation of
additional features that aid the assembly process. Other, entirely different
techniques have also been employed. One \cite{sch98} considered the dynamics of
spherically symmetric particles subject to directional interactions, whose
states and binding energies are selected probabilistically based on rules
involving local neighborhoods \cite{ber94}. Another \cite{bru03} studied disk
packings on a spherical surface using a mean-field statistical mechanical
analysis dependent on curvature and coverage.

No solvent is included in the simulation, an approximation borrowed from
protein folding; even a neutral solvent would increase the computational effort
substantially without influencing the outcome, not only because of the
additional particles involved, but also because of the slower capsomer dynamics
when moving in a solvent rather than in a vacuum. (In future detailed work,
accounting for the effect of, e.g., pH and salt concentration on the pathways
and the final state, will require more substantial models that address
interfacial interactions and solvation effects at a molecular level. An
alternative, Langevin-like representation of the solvent contribution as small
random impulses, suitably distributed over space and time, has also not been
used; if the dynamics are dominated by intermolecular forces, as is the case
here, there will be little overall effect on the eventual outcome.)

The need to assure defect-free assembly allows a design tradeoff between model
capsomer size and interaction complexity. Bonding of real capsomers involves
the participation of a relatively large number of interaction-site pairs across
a mutual contact surface. Although such interactions are not individually
directional, in view of the substantial size of the capsomer compared to the
effective interaction range, it is unlikely that the interactions are capable
of producing strongly bound states in which capsomers are incorrectly aligned.
This is not necessarily true for simplified models with few interaction sites
and an overall capsomer size similar to the range of attraction, where small
clusters can become trapped in states corresponding to spurious local energy
minima instead of developing into complete shells.

One method of avoiding this situation was introduced for small capsomer models
and employs rules governing the interactions (Section~\ref{sec:rules}) in
addition to permanent bonds \cite{rap99}. The alternative is to use an enlarged
capsomer with more interaction sites and reversible bonds; the computational
cost is increased, but the enhanced structural rigidity reduces the opportunity
for incorrect bonding. These design alternatives demonstrate possible
approaches to the problem. They also reflect a changing perspective brought
about by the fact that the available computer resources continued to increase
throughout the study; greater computing power allows better capsomer
representations, larger systems and longer runs, all contributing to the
eventual shift from permanent to reversible bonding.

\subsection{\label{sec:struct}Capsomer structure}

The simplified trapezoidal capsomer representation employs a rigid assembly of
several soft (more precisely, almost hard) spherical particles arranged in a
partly overlapping configuration approximating the desired shape; specific
designs are described in Section~\ref{sec:specdes} and illustrated in
Figures~\ref{fig:01}--\ref{fig:03} below. The repulsive force between spheres
used to prevent significant spatial overlap is based on the truncated
Lennard-Jones potential
\begin{equation}
u(r) = \left\{
\begin{array}{ll}
4 \epsilon [(\sigma / r)^{12} - (\sigma / r)^{6} + 1/4] & \quad r < r_c \\[4pt]
0 & \quad r \ge r_c \label{eq:ss}
\end{array}
\right.
\end{equation}
where $r$ is the sphere separation, $\sigma$ approximates the sphere diameter,
$\epsilon$ determines the energy scale, and $r_c = 2^{1/6} \sigma$ is the
interaction cutoff; in the reduced units used subsequently, $\sigma = \epsilon
= 1$. A suitably arranged set of spheres provides an adequate approximation to
the desired shape, but as more spheres are used the computational effort
required to evaluate the interactions between nearby molecules grows. The
description based on pair interactions alone is, however, much simpler than
alternative shape representations that require evaluating the overlap of
complex rigid bodies.

Capsid size limits the amount of genetic material that can be stored. Capsids
consisting of 60 capsomer units exist, but their internal volumes are
insufficient for substantial amounts of nucleic acid; shells are generally
larger, consisting of multiples of 60, i.e., 60\,T, units, where the
triangulation number T can have values from 1 to beyond 25. All sizes share the
icosahedral symmetry, but since 60 is the maximum size under conditions of
complete equivalence, the concept of quasi-equivalence is invoked to explain
how larger structures can be constructed \cite{cas62,cas85,ros85}. Inspiration
for this generalization came from the geometrical principles developed by
Buckminster Fuller for geodesic domes, and construction entails arranging 12
pentamers and certain specific numbers of hexamers in a symmetric manner to
produce a close approximation of a sphere that retains the 60-fold symmetry. If
identical capsomer proteins are used, but a small amount of conformational
deformation permitted (remaining within acceptable variations for bond lengths
and angles) to achieve a minimum free energy structure, the same design
principles lead to a general way of constructing shells, with the ubiquitous
icosahedral symmetry as a necessary consequence.

Viewed from a simplified geometric perspective, hexamers are planar oligomers
constructed out of six triangular triplets, each consisting of three
trapezoidal capsomers. A nonplanar pentamer can be formed by removing one
triangle from the hexamer and closing the gap. In the assembly of a polyhedral
shell, the natural tendency to grow hexagons at certain positions is overcome
by the more global free-energy benefits of forming a pentagon. Molecular
`switches', a conformation-modifying mechanism known as autostery, regulate
formation of hexamers or pentamers, while ensuring these different
subassemblies are positioned appropriately in the surface lattice. Without this
conformational polymorphism, assembly would produce either flat hexamer sheets
or T=1 polyhedra formed entirely of pentamers. In the case of
quasi-equivalence, autostery represents an important characteristic of the
process as it allows otherwise identical capsomers to occupy spatially
nonequivalent locations in the shell. (Quasi-equivalence is only a general
design consideration, and while appropriate for some viruses, may offer only a
partial explanation for others; a full understanding of any given capsid
structure requires analysis of the bonding energies of the capsomers
themselves.)

Implicit in the model design is the fixed shape. This issue is avoided when
studying T=3 assembly by defining three slightly different capsomer shapes
(Section~\ref{sec:specdes}) destined to occupy different classes of locations
within the shell. Explicit inclusion of an autosteric mechanism to provide the
conformational changes required by quasi-equivalence \cite{red98} -- mechanical
analogs of such processes are discussed in \cite{cas80} -- would complicate the
models.

The capsomer bonding forces are associated with interaction sites suitably
positioned within the simplified structures, as shown in
Figures~\ref{fig:01}--\ref{fig:03} below; specific pairs of sites interact
whenever they approach to within a given range, and the closed-shell
configuration corresponds to the minimum-energy state in which adjoining
capsomers are correctly oriented; bond interactions are discussed in
Section~\ref{sec:bondint}. In a properly implemented design the only structures
that should be capable of self-assembly are partial or complete shells, with
energetic considerations excluding an enormous variety of `mutant' structures.

\subsection{\label{sec:specdes}Specific designs}

The smaller of the two cases considered here is a capsid shell of 60 identical
trapezoidal capsomers. The shell can be regarded as an icosahedron \cite{wil79}
each of whose 20 equilateral triangular faces is subdivided into three coplanar
trapezoidal units representing the capsomers of a T=1 virus, as shown in
Figure~\ref{fig:04} below. (Earlier work \cite{rap99} dealt with the simpler
task of assembling 60-faced pentakisdodecahedra from triangular units.) The
lateral capsomer faces within the triangle are normal to the triangular plane,
whereas those along the outside of the triangle are inclined at $20.905^\circ$
to the normal, resulting in a dihedral angle of $138.190^\circ$. Successful
assembly is conditional upon correct relative dimensions to ensure components
fit together, and angles consistent with the overall shell curvature.

The larger case corresponds to a T=3 virus. The capsid is based on a rhombic
triacontahedron \cite{wil79} with 30 identical rhombic faces; each face is
subdivided into two isosceles triangles (the base angles are $58.283^\circ$, so
the triangles are almost equilateral), and each of these triangles is then
divided into three coplanar trapezoidal capsomers yielding a total of 180; the
assembled shell appears in Figure~\ref{fig:05} below. The lateral faces within
the same triangle, and between the triangles comprising the rhombus, are normal
to the triangular plane, while the other faces are inclined at $18^\circ$,
producing a dihedral angle of $144^\circ$. As discussed in
Section~\ref{sec:struct}, the use of three capsomer variants with slightly
different face angles (and attractive interactions only between corresponding
face pairs) avoids the quasi-equivalence issue.

\begin{figure}
\includegraphics[scale=2.0]{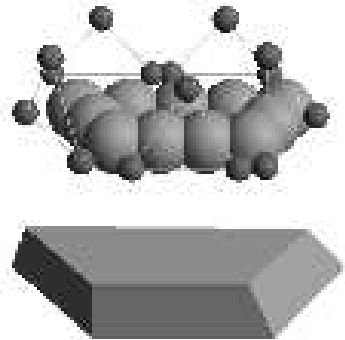}
\caption{\label{fig:01} (Color online) Capsomer model used for T=1 simulation
with permanent bonding; the spheres and interaction sites (small spheres)
comprising the capsomer and the effective trapezoidal shape are shown.}
\end{figure}

Figure~\ref{fig:01} shows the trapezoidal capsomer used in the T=1 permanent
bond studies. The larger, slightly overlapping spheres that provide the overall
shape occupy a single plane, while the interaction sites, represented by small
spheres for visual convenience, determine the locations and orientations of the
lateral faces. Each bond requires the coupling of three complementary pairs of
interaction sites. The edges have between one and three interior spheres, and
the triangular sets of bonding sites extend well beyond the volume of the
capsomer spheres (pointing in the direction of the shell interior when the
capsomer is correctly positioned). Two representations are shown; one in terms
of the spheres and interaction sites, the other a block approximating the
overall shape; this simplified model should be contrasted with real capsomers
\cite{bak99} consisting of long, folded proteins whose exposed surfaces form
relatively complex landscapes.

\begin{figure}
\includegraphics[scale=1.7]{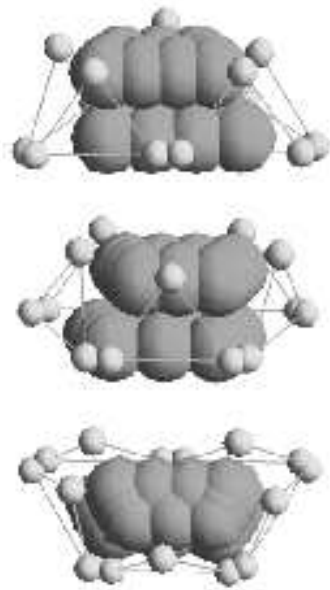}
\caption{\label{fig:02} (Color online) Views of T=3 capsomer model.}
\end{figure}

The corresponding T=3 version is shown in Figure~\ref{fig:02}. Since the three
capsomer types differ only slightly, just one is shown, in different
orientations. It has a smaller planar area than for T=1 in order to allow
smaller shells relative to the size of the simulation region, but consists of
two touching layers of spheres providing greater depth that helps avoid
unwanted interactions. The spheres overlap within each layer, and this is
varied to adjust the lateral face slopes.

\begin{figure*}
\includegraphics[scale=1.8]{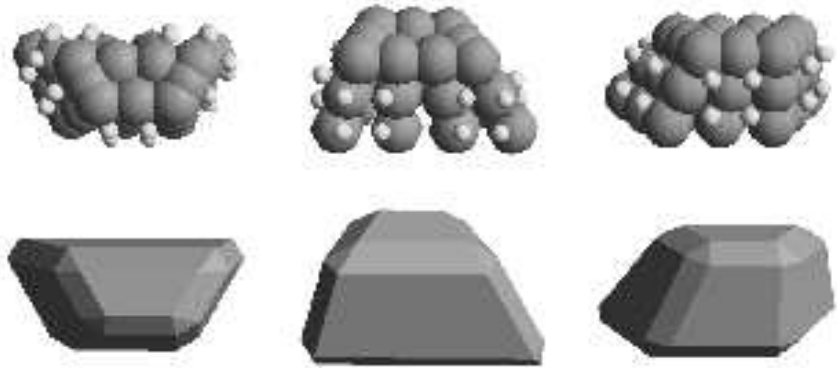}
\caption{\label{fig:03} (Color online) Capsomer model used for reversible
bonding (T=1); different views of the sphere-based structure and its effective
shape are shown.}
\end{figure*}

The final capsomer form, appearing in Figure~\ref{fig:03}, is used for the
reversible bond case with T=1. The shape is produced using three layers of
spheres to reduce even further the likelihood of incorrect bonding; this is now
a more important issue since there are no rules (see Section~\ref{sec:rules})
to help avoid interactions that do not contribute to the final shell. Sphere
spacing within layers varies from overlapping to well-separated. Each bond now
involves four pairs of interaction sites (more closely spaced than before); the
energetic gain of a correctly aligned state is enhanced by distributing the
interactions over more site pairs. Because of the increased capsomer thickness
(three sphere layers rather than one or two) interaction sites can be
positioned without extending beyond the actual area of the lateral faces, and
the resulting steric screening helps prevent unwanted interactions.

Figures~\ref{fig:04} and \ref{fig:05} show complete shells that these capsomers
are capable of forming. (The actual size of these shells can be deduced from
the ratio of icosahedron edge length to radius, namely $\sqrt{5} - 1 =
1.236\dots$.) The former shows a T=1 shell; to make the bond locations visible
the capsomers have been reduced slightly in size. The latter shows a T=3 shell
formed from 180 trapezoidal units of the three different types; here capsomers
are drawn to show their effective sizes hiding the bonds. Due to the nature of
the bonding forces -- see Eq.~(\ref{eq:attr}) -- mutually attracting
interaction sites are spatially coincident in the ground state.

\begin{figure}
\includegraphics[scale=1.3]{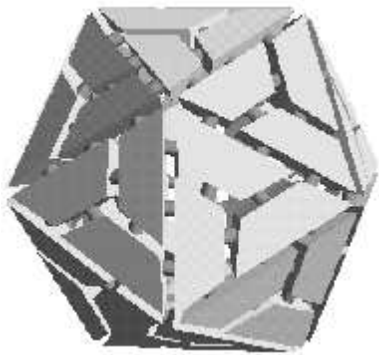}
\caption{\label{fig:04} (Color online) Complete T=1 shell (as produced by the
simulation) with 60 capsomers; capsomers are shown reduced in size so that
bonds are visible.}
\end{figure}

\begin{figure}
\includegraphics[scale=1.3]{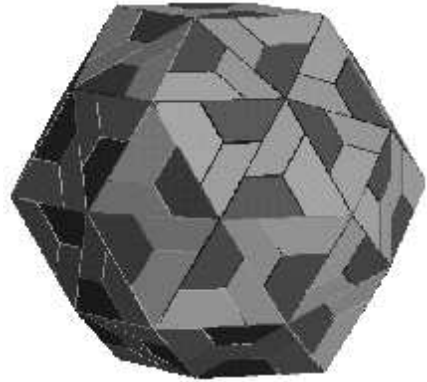}
\caption{\label{fig:05} (Color online) Complete T=3 shell shell with 180
capsomers of three (color-coded) types.}
\end{figure}

The designs leave ample scope for enhancement. More complex capsomer surfaces,
for example, would allow the introduction of a `lock-and-key' mechanism due to
steric effects. An attempt was made to use such a technique in early work on
triangular units, by adding an extra sphere to the center of a lateral face and
leaving an opening in the complementary face, but this mechanism did not
provide the desired additional rigidity; it was not tested with larger units
however, since the increased size and multiple interaction sites accomplish the
same goal of restricting internal degrees of freedom.

\begin{figure*}
\includegraphics[scale=0.32]{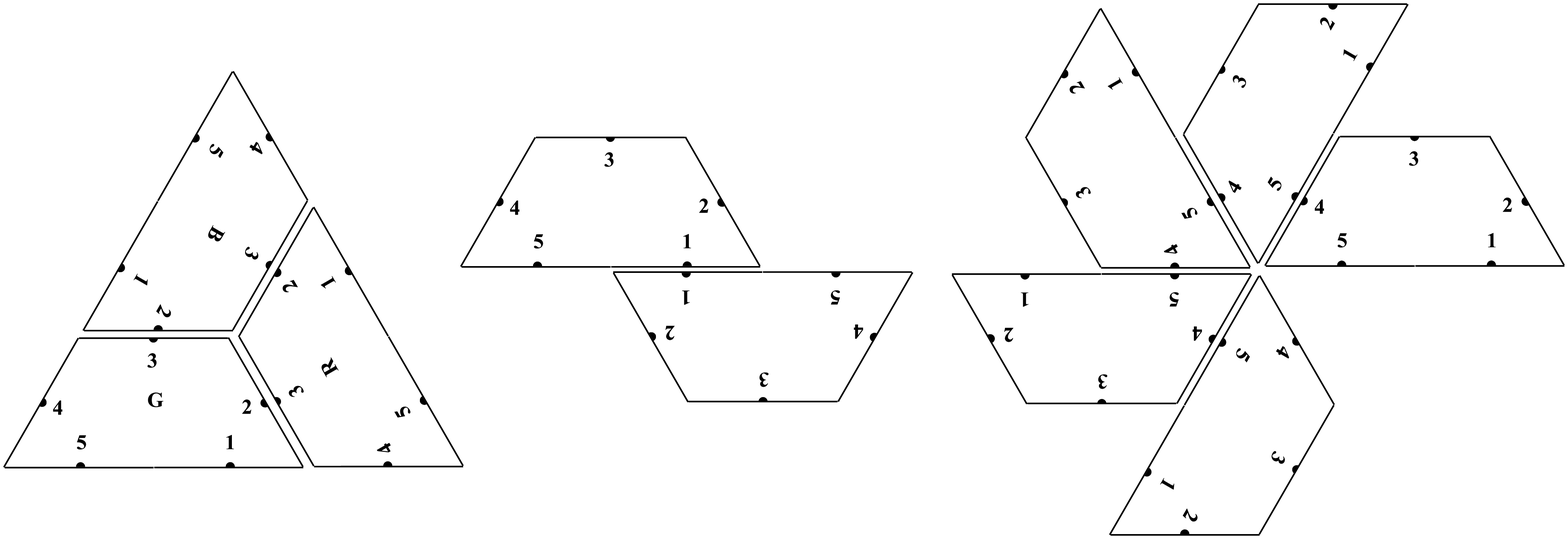}
\caption{\label{fig:06} Dimer, trimer and (opened-up) pentamer configurations
with labeled interaction sites; the color identification (for T=3) is included
in the triangular configuration.}
\end{figure*}

\section{\label{sec:bondint}Bond interactions}

Two fully-bonded capsomers are held together by interactions between sets of
either three or four pairs of complementary interaction sites that can only
interact with one another. The use of multiple sites, in addition to being more
realistic, helps accomplish several goals: (a) The orientation of a lateral
face and, consequently, the dihedral angle between capsomers, is specified by
the plane containing the sites. (b) The overall binding interaction is
distributed over the contact face; this ensures that the total binding energy
of misaligned capsomers can only be a fraction of the ground-state value,
thereby lessening the stability of the bond. (c) Finally, multiple binding
sites enhance structural rigidity by suppressing internal modes such as
twisting or flapping. Typically (though not always), two capsomers are drawn
together initially by just one of the interaction site pairs, and they then
reorient so the remaining site pairs can participate.

As shown in Figures~\ref{fig:01}--\ref{fig:03}, each of the three short
capsomer faces contains a single set of interaction sites, while the long face
contains two sets. The labeling scheme used for the sets appears in
Figure~\ref{fig:06}. Since color coding is useful for T=3, capsomers are
labeled B, G, R (blue, green, red) following the convention used in
\cite{spe95}. For T=1 the trimer forms an equilateral triangle (as shown),
whereas for T=3 a small change of apex angle makes the triangle isosceles
(Section~\ref{sec:specdes}). The five interaction sites follow the same bonding
pattern in each capsomer, namely sites 2 and 3 can bond, as can 4 and 5, while
sites labeled 1 bond with each other. Also shown are the site pairings
associated with trimer, dimer, and pentamer or hexamer formation. Three
capsomers joined by 2-3 bonds produce a planar triangular trimer. A single 1-1
bond forms a dimer; it is nonplanar for T=1, while for T=3 it is planar when
two type G capsomers are involved and nonplanar in the alternative R-B case.
The 4-5 bonds produce a nonplanar, flower-like pentamer for T=1, while for T=3,
if all capsomers are of type B they produce a pentamer, or a hexamer if
alternating R and G types are involved (in three coplanar pairs).

The functional form of the attractive potential between interaction sites is a
negative power of the site separation $r$, provided the sites are not too
close, but for $r < r_h$ it takes the form of a narrow harmonic well; this form
is chosen for convenience, and when the system is at rest $r = 0$. In the case
of reversible bonding, the force is derived from the potential
\begin{equation}
u(r) = \left\{
\begin{array}{ll}
e (1 / r_a^2 + r^2 / r_h^4 - 2 / r_h^2) & \quad r < r_h \\[4pt]
e (1 / r_a^2 - 1 / r^2) & \quad r_h \le r < r_a \label{eq:attr}
\end{array}
\right.
\end{equation}
with typical parameter values $e = 0.1$, $r_h = 0.3$, and cutoff $r_a = 2$. The
interactions used with permanent bonding are similar, although details differ
slightly; earlier work also included an explicit torsional interaction to
accelerate the bonding process, but this was discarded once it became apparent
that pair interactions alone were sufficient.

\begin{figure}
\includegraphics[scale=1.3]{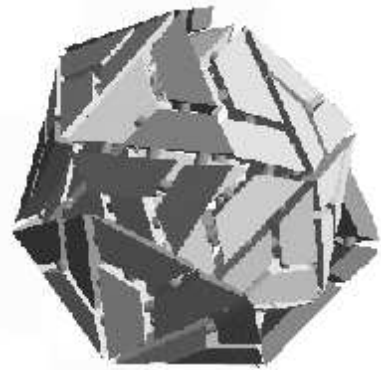}
\caption{\label{fig:07} (Color online) Example of mutant capsid structure
(permanent bonding); capsomer size is reduced, as in Figure~\ref{fig:04}, to
show the bonds (yellow particles are fully bonded, blue are not).}
\end{figure}

At this juncture the two techniques diverge. In the approach used initially,
bond formation is regarded as irreversible, while the alternative is to treat
bonds simply as potential wells of finite depth; these represent, respectively,
the extremes of kinetically limited and equilibrium assembly. Physical
justification for permanent bonds stems from possible conformational changes,
and even cleavage, experienced by proteins in the course of bonding. Formation
of a permanent bond between two interaction sites can be implemented using
Eq.~(\ref{eq:attr}); when the site separation initially falls below $r_h$ the
potential is replaced by the harmonic term alone, irrespective of $r$,
producing an infinitely high barrier from which escape is impossible.
Implementation requires monitoring the identities of interacting site pairs,
but this enables another feature, namely, that once a pair of sites have bonded
permanently they then attract only each other, thereby reducing a tendency to
form amorphous globules.

The rapid assembly associated with permanent bonding implies a more promising
approach than bonds subject to breakage, but exploration reveals certain less
desirable features. Stretchable bonds mean that partially formed assemblies are
subject to structural distortions sufficiently large for bonds to appear
between capsomers that, in more rigid assemblies, would be beyond bonding
range. This leads to defects that not only prevent full assembly but also spawn
mutant structures. Figure~\ref{fig:07} shows a rare defective assembly in which
growth of a second outer layer begins at a misaligned shell element.

This situation is avoidable by making bonds nonpermanent, allowing energetic
considerations to inhibit long-term survival of incorrect bond pairings. The
severity of the problem can also be reduced for permanent bonds, but this
requires the introduction of additional features -- assembly or interaction
`rules' -- some physically motivated, others arbitrary; since similar issues
are likely to appear in other kinds of self-assembly simulations, a brief
discussion of the rule-based approach appears in Section~\ref{sec:rules}.

\section{\label{sec:assem}Assembly scenarios}

\subsection{\label{sec:paths}Pathways}

Polyhedral shells can assemble in many ways, even after allowing for the
icosahedral symmetry. The presence of systematic features in the construction
process is an important issue, since the likelihood of two arbitrary
subassemblies being able to mesh successfully is low (unless incompatible
pieces can be discarded in the process), but whether preferred assembly
pathways exist is unknown. One hypothetical assembly scenario is based on a
multistage process, with small clusters of specified shape forming initially,
and then combining into increasingly larger subassemblies. These small clusters
must be able to `tile' the full shell, so a possible first stage is the
assembly of triangles from the trapezoidal capsomers, while in the second stage
these trimers are added (one at a time) to build full shells (combining
subassemblies each consisting of several trimers is again an event with a low
success rate). Experimental signatures of multistage assembly would be certain
preferred intermediate cluster sizes, or the more accessible rate concentration
dependence \cite{zlo99,zlo00}. In simulations that involve permanent bonding,
assembly order can be enforced by using bonding rules, as described in
Section~\ref{sec:rules}.

In the case of reversible bonding, where bond breakage occurs as a consequence
of thermal fluctuations, assembly order can be biased by energetic preferences;
in order to encourage early dimer or trimer formation, larger force constants
would be associated with the relevant bonds (here, the strength $e$ is
doubled). Left unattended, the system evolves to a state with many partial
shells, from which further growth is impossible; to ensure there is always an
adequate supply of free capsomers, partial shells below a certain threshold
size are broken up at regular intervals (how often depends on growth rate) by
switching off their bonding interactions for a short period. (A possible
modification useful for quantitative work entails breaking up newly completed
shells as well, to increase the number of shell-growth histories available for
study.)

\subsection{\label{sec:rules}Interaction rules}

Permanent bonds do not allow construction errors to be rectified, so it is
crucial to reduce any opportunity for incorrect bond formation; this is
accomplished by introducing rules governing when attractive forces may act.
Analogous rules could appear in real molecules through local conformational
variation in response to changes in bonding state, although their existence is
not readily established; here, rules are introduced to compensate for other
design simplifications. As a historical aside, the simulations started with
small polyhedra, ranging from tetrahedra to 32-faced `soccer balls', with
interaction rules playing a prominent role; rule complexity increased for
larger polyhedra with dubious chemical justification, so the rule-based
approach with permanent bonding was eventually supplanted by larger capsomers
that allow greater freedom in positioning interaction sites and reversible
bonding to reduce construction errors.

The most basic of the rules aims to avoid bond formation in ways inconsistent
with the final structure: Over the time interval starting when one of a
capsomer's interaction sites bonds with the complementary site on another, and
ending when all other sites in the set are joined, these capsomers cannot form
other bonds. If construction follows a pathway in which, for example, trimers
initially form, which then bond into larger structures, an analogous rule
applies to entire trimers. In order to minimize any adverse effects (but
without attempting to mimic nature which deals with much larger systems and is
not necessarily reliant on high yields), if bonding fails to complete within a
prescribed interval the existing partial bond is broken; to prevent immediate
rebonding newly separated capsomers must wait for a certain time before they
can begin bonding again. Such rules help ensure the release of units that
cannot bond completely, as in the case of two capsomers attempting to bond
simultaneously along different edges of a shell opening big enough for only
one.

Other interaction rules enforce assembly pathways (Section~\ref{sec:paths});
thus if growth occurs via trimer intermediates, capsomers are first required to
form trimers, and only when all internal bonds are complete can these clusters
associate into larger structures. Additionally, by restricting the number of
larger subassemblies that can nucleate by the joining of (e.g.) two trimers --
equivalent to a rate-limiting process \cite{zlo99} -- it is possible to ensure
a significant yield of complete shells rather than numerous fragments.
Typically, this limit would be set so that 50--75\% of the trimers are used by
the full shells, while the rest supply an adequate background concentration of
construction material; a similar restriction is applied to the formation of the
trimers.

Because of the intrinsic bond flexibility, the need for further rules only
becomes apparent as incorrect structures are encountered; two examples will be
mentioned here. When the pathway involves dimers, the appearance of unwanted
bonds is reduced by requiring that after a 1-1 bond creates a dimer, the next
must be a 2-3 bond between one of the dimer members and a capsomer in a growing
shell, and only then the remaining 4-5 bond; since the 2-3 bond joins sites
closer to the dimer center than the 4-5 bond, this restricts partially bonded
units from encountering inappropriate bonding partners. For trimer assembly,
the 2-3 bonds are used to build the trimer, the 1-1 bond may then join the
trimer to a shell and, last of all, the more distant (from the trimer center)
4-5 bonds are allowed to form. Such constrained bonding sequences might be
attributable to conformational changes as bonding progresses.

\section{\label{sec:tech}Simulation techniques}

General MD methodology is discussed in \cite{rap04}; here a brief summary of
the issues relevant to capsomer simulation will suffice. Interaction
calculations are carried out efficiently using neighbor lists; list
construction follows the procedure used for monatomic fluids. Separate lists
are used for short-range repulsive forces between the spheres giving capsomers
their shape, Eq.~(\ref{eq:ss}), and for longer-range attractive forces between
interaction sites, Eq.~(\ref{eq:attr}). The rotational equations of motion
employ standard rigid-body methods; these, together with the translational
equations, are solved using a leapfrog integrator. The simulation region is
bounded by elastically reflecting hard walls, implemented using short-range
repulsive forces based on Eq.~(\ref{eq:ss}) acting normal to the surfaces;
since visualization is important, hard walls avoid potentially confusing
imagery accompanying periodic boundaries. In the initial state, units are
positioned on a lattice and assigned random orientations and velocities;
lattice spacing determines the mean density of the capsomer `fluid'.
Simulations based on reversible bonding -- which avoid the complexity
associated with interaction rules -- can be run on a distributed memory
(message passing) parallel computer for improved performance.

Exothermal bond formation gradually heats the system; this problem is
particularly acute due to the limited number of degrees of freedom capable of
absorbing the excess energy (the reason being the absence of solvent and the
rigid capsomer structure). Applying a weak damping force $- \gamma (\vec{v}
\cdot \vec{r}) \vec{r} / r^2 $ along each bond resolves this issue, where
$\vec{v}$ is the relative velocity of the interaction sites and the damping
coefficient $\gamma = 0.1$. Use of constant-temperature MD ensures that the
overall temperature does not change despite bond formation and damping; the net
effect is to transfer energy associated with internal vibration to the motion
of entire clusters.

The interaction parameters are chosen for efficient self-assembly while
maintaining numerical stability; there is presently no relation to experimental
association energies \cite{red98}. The number density affects the outcome and
must also be established empirically: Too high a value will not provide
adequate space for shells to grow without mutual interference; a density that
is too low will retard growth due to capsomers lying beyond their attraction
range and the lack of collisions that can, in the case of reversible bonding,
help break off incorrectly bound units from partially constructed shells (the
densities actually used are substantially higher than in experiment).

In the case of permanent bonding, the T=1 simulations employ 1000 capsomers,
with 13 shells allowed to nucleate and run lengths of 2--300,000 steps; for the
larger T=3 shells the system contains 4096 capsomers, 10 shells can nucleate
and run lengths are 5--800,000 steps. Bonds are allocated 4000 steps to form,
and if unsuccessful the participating sites have their attractive forces turned
off over the next 1000.

The reversible bonding runs for T=1 follow 4096 capsomers over 10 million
steps, with no limit on shell nucleation. Clusters with size $\le 30$ are
broken up every 500,000 steps by turning off their attractive interactions over
the subsequent 10,000 steps. The existence of a bond (insofar as cluster
measurements in Section~\ref{sec:res} are concerned) depends on the separation
of either a single pair of interaction sites or -- the more stringent
requirement -- on the separations of all (either three or four) pairs
associated with the bond; in either case the interaction sites must approach to
within a distance 0.2 ($< r_h$) to be considered paired. Finally, in reduced MD
units, capsomers have unit mass, the integration step is 0.005, the temperature
unity, and the number density typically 0.004.

\section{\label{sec:res}Results}

\subsection{Shell analysis}

The simulation results are both quantitative and qualitative in nature; a
`snapshot' sequence recorded at intervals of 2000 steps over the course of each
simulation run provides the data needed to recreate capsid growth for
post-analysis. A run can require several days of computing on a powerful
workstation, but subsequent processing requires only minimal effort. Shell
properties are readily measured, allowing the mean growth statistics to be
analyzed, together with the behavior of individual shells (time resolution is
limited by the snapshot interval, so shortlived bonds between snapshots are
missed). Animated sequences providing a condensed summary of the run, in full
three-dimensional detail, are also available, although static images will have
to suffice here.

Establishing shell completeness requires (a) identifying a bonded set of
capsomers using cluster analysis \cite{rap04} with the appropriate bonding
criterion, (b) ensuring their number equals the expected shell size, and
finally (c) checking that each capsomer is bound to the correct number of
neighbors; if all these tests succeed then, in view of the comparative rigidity
of the bonds, the cluster corresponds to a closed shell. It is somewhat
arbitrary whether all site pairs, or just a single pair, must be within range;
the two criteria tend to complement one another. The former, more stringent
criterion produces smaller clusters during early growth; this is helpful when
trying to visualize the emergence of partial shells from the capsomer `soup',
but omits the local environment to which a growing cluster might be be loosely
attached. The latter produces larger clusters, including big, ill-defined
structures that incorporate multiple partial shells. During later stages of
assembly, both criteria lead to similar results, since as shells near
completion correct capsomer positioning allows all sites to participate in
bonding.

Describing the nature of incomplete shells, while straightforward when the
imagery is available (a nearly complete shell is readily characterized, as is a
shell with a localized defect), is not obviously quantifiable; since partially
formed structures, even defect-free shell fragments, have a variety of
morphologies, mechanizing their classification is nontrivial. Each such
structure can be represented as a bonded network (or graph), and while it is
possible to determine the topology by evaluating connectivity, and the
compactness by counting missing bonds, it is not apparent how such information
can be utilized. Furthermore, development is not necessarily a process whereby
shells grow monotonically by accretion of individual capsomers or small
subassemblies, since it is also possible for larger subassemblies to aggregate
and (with reversible bonding) for groups forming partial shells to break away
from larger structures.

\subsection{Growth}

The most important observation regarding the overall behavior, equally
applicable to both permanent and reversible bonding scenarios, is that
polyhedral shells have little difficulty growing to completion, and mutant
structures are highly unlikely. Partial shells tend to have few voids in their
surfaces, and shells nearing completion typically have only one or two holes;
more open cagelike structures with multiple lacunae are not encountered. The
results also reveal, not surprisingly, that an inappropriate choice of
interactions (or, for that matter, even a slight error in defining capsomer
geometry) leads to a wide variety of alternative structures, including open
networks, incorrectly linked assemblies of shell fragments and amorphous shapes
that defy characterization. Due to the difficulty of describing anything other
than a correct shell, this aspect of the subject is avoided, but real viruses
experience analogous effects in unfavorable environments.

\begin{figure}
\includegraphics[scale=0.75]{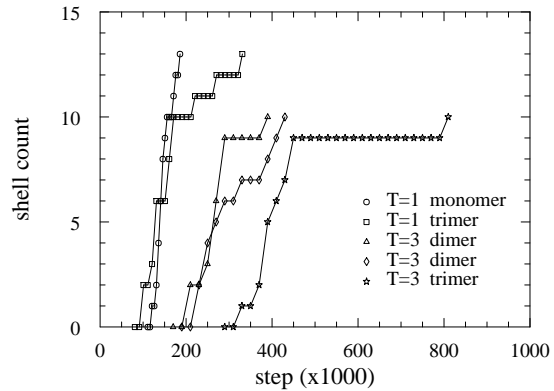}
\caption{\label{fig:08} Number of complete shells vs time for permanent
bonding.}
\end{figure}

\begin{figure*}
\includegraphics[scale=1.6]{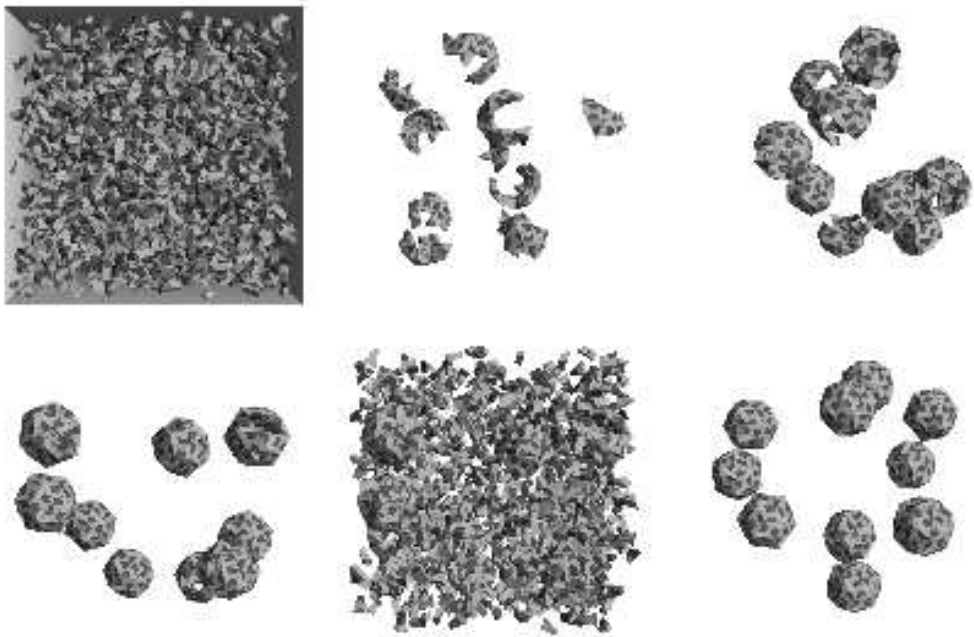}
\caption{\label{fig:09} (Color online) Snapshots from a T=3 simulation with
permanent bonding: the images show an early state (including the container for
reference); three views at different times showing only the growing shells; the
entire system corresponding to the third of these views; the final state
(capsomers are colored by type).}
\end{figure*}

The main emphasis of the analysis is on reversible bonding, the likely focus of
future work, but a few key results for the permanent bonding alternative are
presented first in order to demonstrate its capability. Figure~\ref{fig:08}
summarizes the number of complete shells as a function of time for several
different cases, namely T=1 shells grown using monomer and trimer pathways, and
T=3 shells using dimer and trimer pathways, with two examples of the former to
show the variation between runs. Figure~\ref{fig:09} shows several snapshots of
T=3 growth using a trimer pathway; shells are isolated from their milieu using
automated cluster analysis.

\begin{figure}
\includegraphics[scale=0.75]{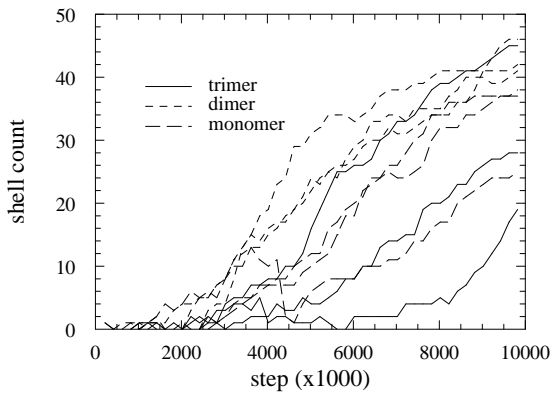}
\caption{\label{fig:10} Number of complete shells vs time for reversible
bonding.}
\end{figure}

\begin{figure*}
\includegraphics[scale=1.6]{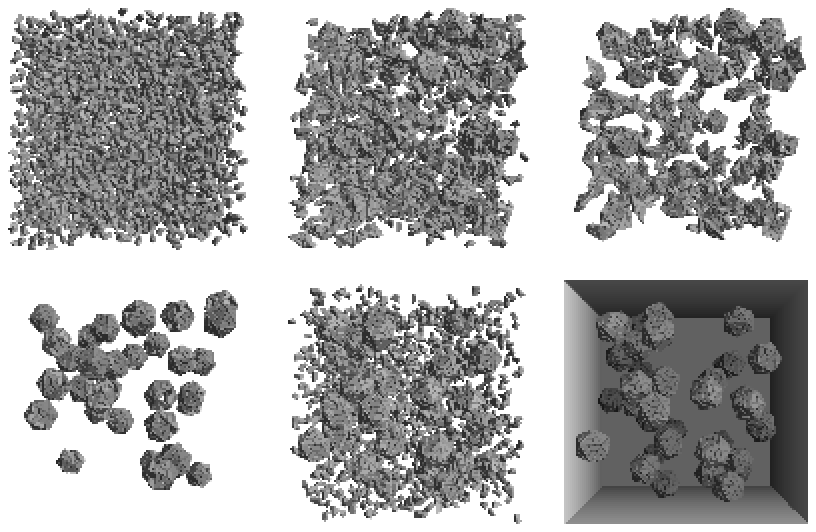}
\caption{\label{fig:11} (Color online) Snapshots from a simulation with
reversible bonding: early state; two views after $0.5 \times 10^6$ steps
showing the full system and the 132 clusters of size $\ge 10$ (color coding -
red particles have all five bonds in place, blue have $< 5$); the 34 clusters
of size $\ge 50$ after $2.5 \times 10^6$ steps; two views of the final state
after $8 \times 10^6$ steps showing the entire system, and just the 39 complete
shells with container.}
\end{figure*}

Figure~\ref{fig:10} summarizes the yield of complete T=1 shells vs time for
simulations employing reversible bonding, three runs each for unweighted,
dimer- and trimer-weighted pathways; note that the runs are now an order of
magnitude longer. Dimer growth appears to produce the highest yields at
intermediate times, but any observation of this kind is tentative because the
spread in trimer results over the runs suggests more sampling is required and
also because the effectiveness of selectively increased interaction strengths
in biasing the pathways has yet to be established. Growth snapshots using
trimer weighting appear in Figure~\ref{fig:11}.

\subsection{\label{sec:stats}Shell statistics}

\begin{figure}
\includegraphics[scale=0.75]{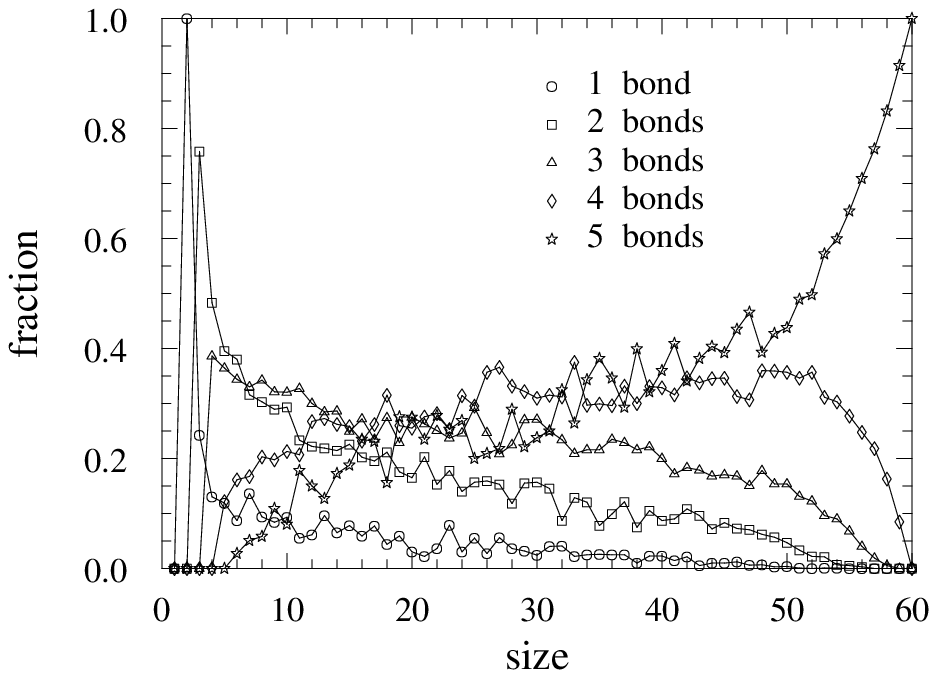}
\caption{\label{fig:12} Capsomer fractions with different bond counts vs
cluster size for reversible bonding and unweighted assembly.}
\end{figure}

\begin{figure}
\includegraphics[scale=0.75]{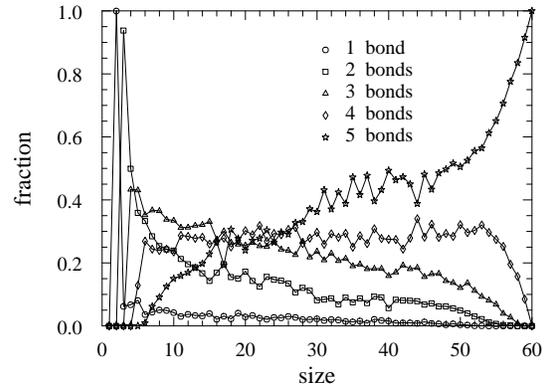}
\caption{\label{fig:13} As in Figure~\ref{fig:12}, for trimer-weighted
assembly.}
\end{figure}

Despite the difficulty in quantifying shell growth, average properties of
partial shells can be used in studying growth trends. Capsomers belonging to
incomplete shells can be categorized according to the degree of bonding; a
higher number with the maximal five bonds is consistent with a more compact
hole-free partial shell, whereas an increased number with just one or two bonds
is an indication that at least part of the structure is loosely connected.
Figures~\ref{fig:12} and \ref{fig:13} show the capsomer fractions as functions
of shell size for unweighted and trimer-weighted assembly; averaging is over
the entire run and the stricter (all-pairs) definition of bonding is used. Two
differences are apparent; in the unweighted case there is a larger proportion
of singly bonded units in the small (size $< 20$) clusters, while in the trimer
case there is an enhanced preference for five bonds over four in the size range
30--50 (possibly reflecting a drop in monomer-sized holes in the shell); both
of these are indicators that trimer weighting leads to more compact
intermediate states (dimer results are similar).

\begin{figure}
\includegraphics[scale=0.5]{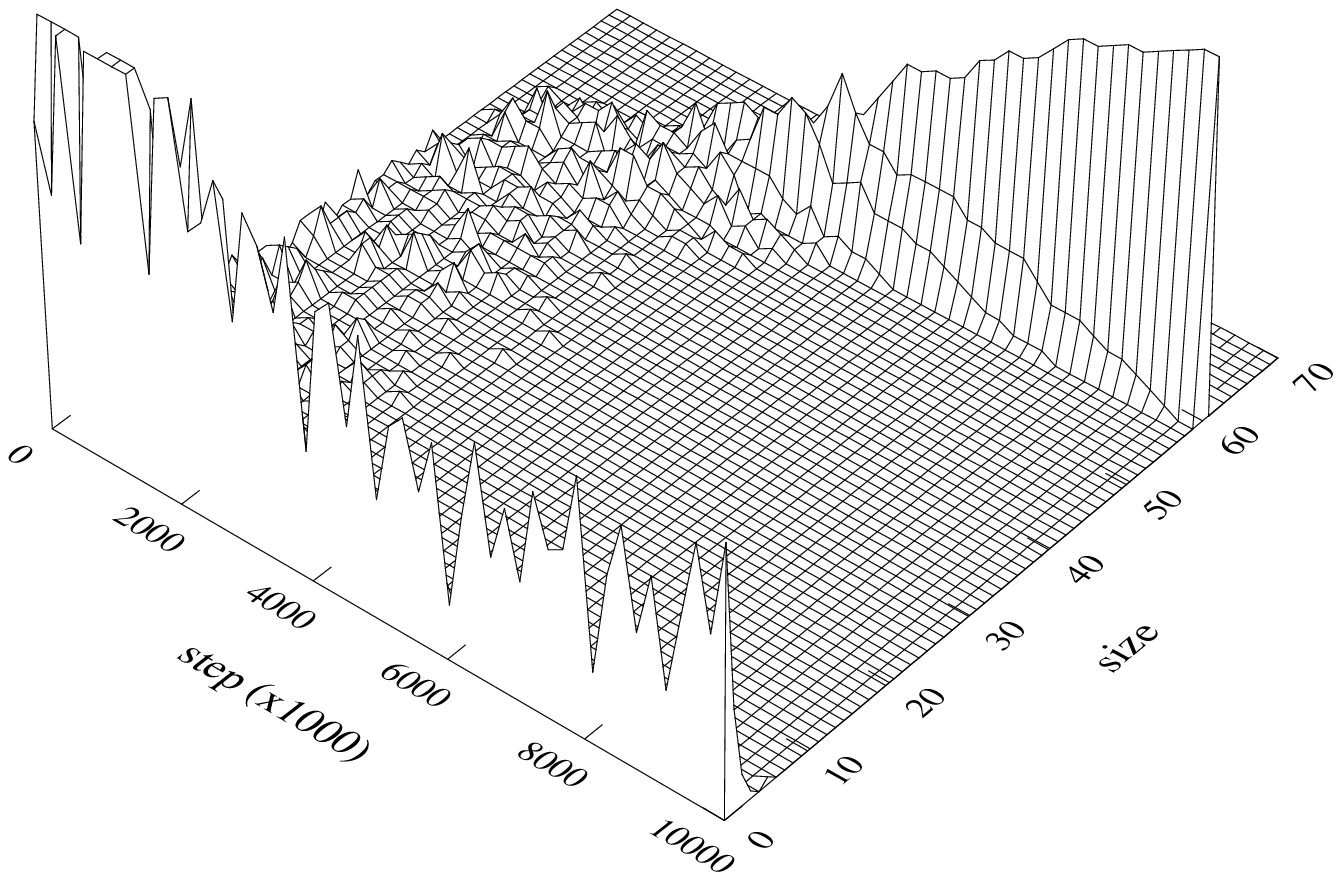}
\caption{\label{fig:14} Cluster size distribution vs time for the
trimer-weighted case.}
\end{figure}

Figure~\ref{fig:14} shows how the size distribution develops with time;
prominent features are the steady trend to completion, a relatively narrow
distribution once growth is well underway, and the imposed cluster breakup. The
fact that for most of the run there are few clusters of intermediate size is a
consequence of breakup followed by prompt attachment of newly freed units to
larger assemblies. At the end of the run the system consists entirely of shells
that are either complete, or nearly so, together with monomers and small
fragments.

\begin{figure}
\includegraphics[scale=0.75]{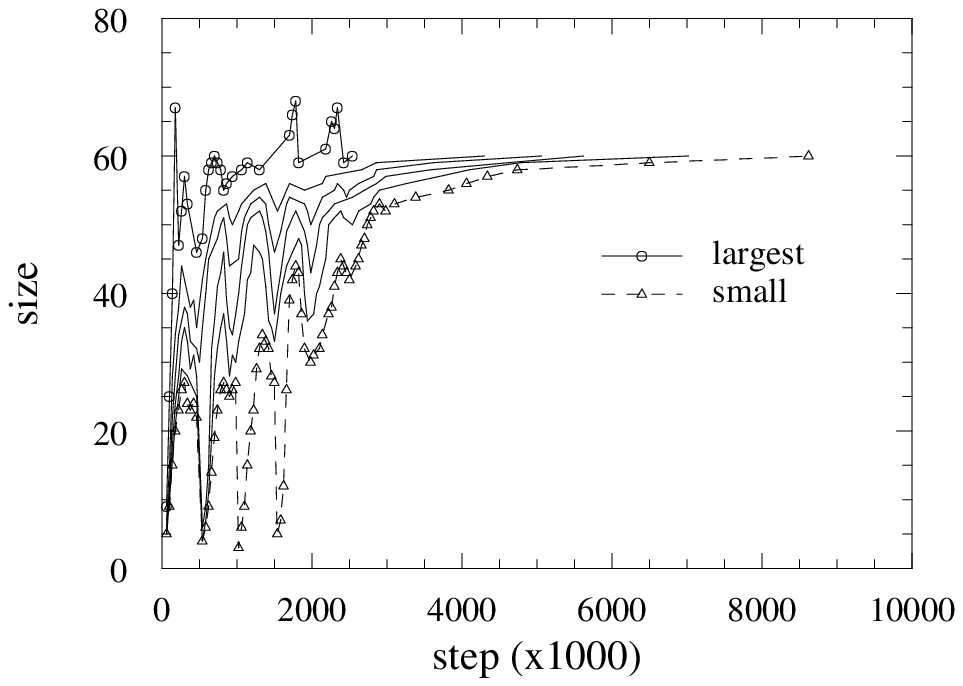}
\caption{\label{fig:15} Ranked cluster sizes for trimer-weighted assembly (only
every 5th cluster is included); all site pairs are required for bonding.}
\end{figure}

\begin{figure}
\includegraphics[scale=0.75]{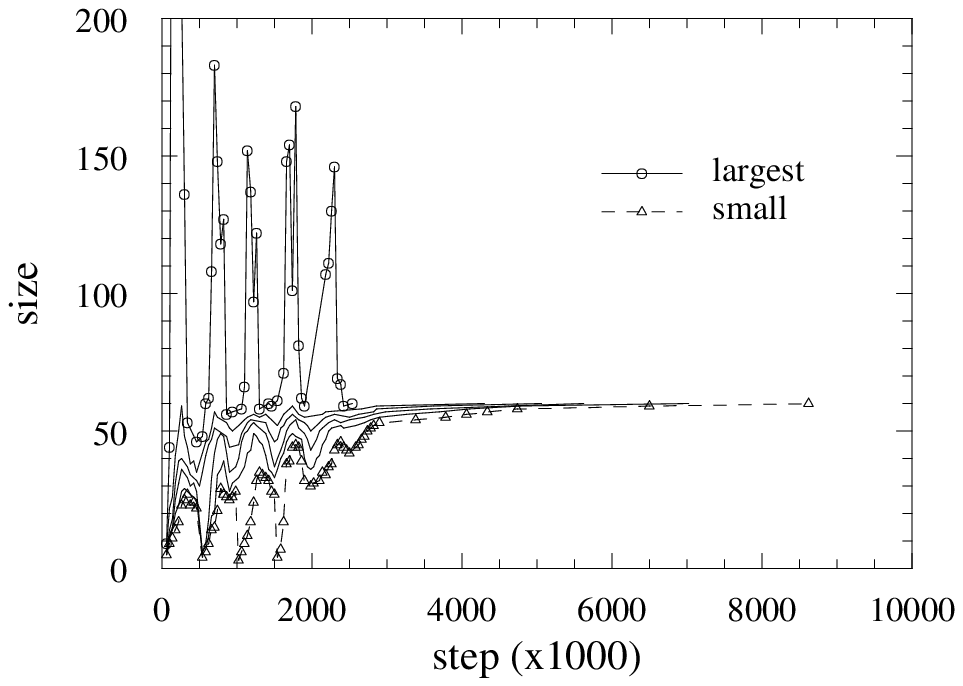}
\caption{\label{fig:16} As in Figure~\ref{fig:15}, but bonding requires only a
single site pair.}
\end{figure}

Detailed information extending beyond such system-wide averages is obtained by
examining individual shell growth; two forms of analysis are demonstrated here.
The first ranks the clusters in each snapshot configuration by size and then
graphs the size history of the largest of these clusters, the 5th largest, and
so on, until the count reaches the number of shells that grow to completion; no
attempt is made to ensure `continuity' by tracking specific clusters. The
results, using the strict all-pairs bond definition, appear in
Figure~\ref{fig:15}; Figure~\ref{fig:16} shows the corresponding results when
only a single pairing is required, leading to larger clusters early in the run.
The long-term behavior in both cases is the same (graphs terminate when shells
are complete). The periodic breakup affects small clusters until their size
exceeds the threshold; there is a secondary effect on larger clusters since
newly freed capsomers provide competition as bonding partners.

The second kind of analysis tracks, as closely as possible, the development of
particular shells. This is accomplished by first identifying all complete
shells in the system at the end of the run, together with their constituent
units, and then using this information while following the construction history
of each shell. It is optional whether to include capsomers whose shell
membership is transient, an effect that appears early in assembly, but less so
later on when a high degree of bonding makes escape more difficult. Ambiguity
can arise as to a shell's true ancestry; where there are several contributing
precursor assemblies, it is the one most heavily represented in the final shell
that is credited.

\begin{figure*}
\includegraphics[scale=0.9]{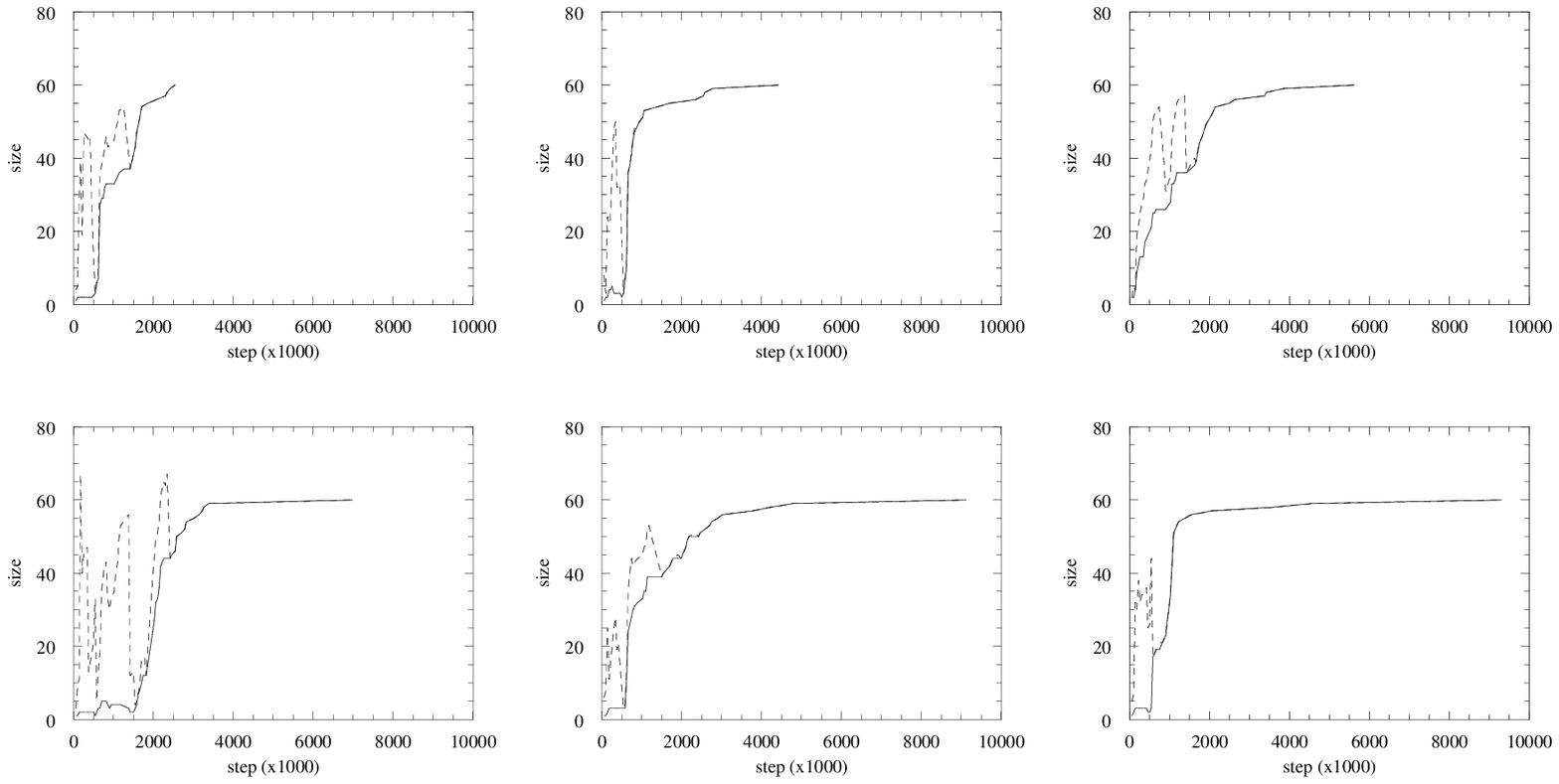}
\caption{\label{fig:17} Size histories of selected clusters that grow into
complete shells (see Section~\ref{sec:stats}).}
\end{figure*}

Figure~\ref{fig:17} shows typical shell growth histories for a trimer-weighted
pathway. The selected shells demonstrate different growth characteristics; they
include both the fastest shell to complete, and one of slowest, taking some
three times longer, although in most cases a delay beyond about two million
steps is caused by a wait for the final few capsomers (possibly just one) to
bond. The strict bond definition is used and histories end upon shell
completion; the almost monotonically increasing solid graphs show how many
members of the final shell have already joined, while the more variable dashed
graphs include transient members that eventually break away from the shell.

Visualization is extremely useful for probing the details of the behavior;
unfortunately its rich message cannot be transferred to the printed page.
Suitable color coding, based on known final shell membership, allows the
identification of units destined to join a particular shell, as well as
transient members. Capsomers, literally, can come and go; only when a capsomer
is bound to most of its neighbors, and embedded in a substantial shell
fraction, is it unlikely to be knocked out of position. Population exchanges of
this kind are not readily characterized in a quantitative manner.

\section{\label{sec:other}Other self-assembly examples}

Although the focus of the paper is on polyhedral shell growth, with the goal of
modeling capsid formation, the same approach works for other self-assembling
systems. Two examples of this kind will be mentioned briefly, the first broadly
related to micelles, the second a unique and somewhat improbable structure that
solves a well-known assembly puzzle. While the former consists of large number
of identical components, the latter has highly specific interactions between a
small set of distinct components. The motivation underlying these examples is
to show what can be accomplished in the simplest of self-assembly simulations.
Related `analog' simulations have been performed in the laboratory, using
millimeter-size plastic objects in solution with appropriate adhesive-coated
surfaces \cite{bre99}.

\begin{figure}
\includegraphics[scale=1.7]{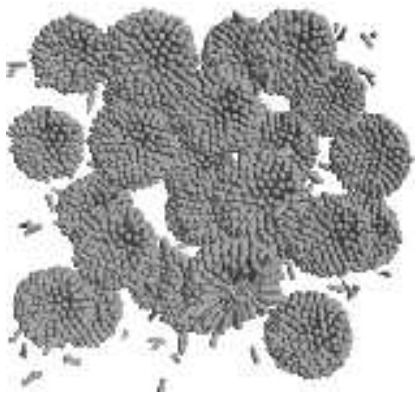}
\caption{\label{fig:18} (Color online) Tapered cylindrical particles that have
condensed into spherical clusters.}
\end{figure}

The first of the systems is a fluid of rigid particles (micelle simulations
\cite{ess94} generally use flexible chain molecules) each having the form of a
tapered cylinder made from a linear array of spheres of decreasing size.
Lennard-Jones interactions occur between spheres occupying equivalent locations
in different particles. The final state of an 8000-particle simulation under
conditions of gradually reducing temperature (a way of encouraging collapse
into the ground state) is shown in Figure~\ref{fig:18}; it consists of 28
packed spherical clusters with sizes ranging from a maximum of 368 down to 260,
one hemispherical cluster of size 183 (half the maximum size), and a few
monomers and tiny clusters. The range of cluster sizes is determined by how
many particles of a given shape can form a packed shell, with the absence of
well-characterized bonding patterns accounting for the size variability, just
the opposite of what occurs with polyhedral shells.

\begin{figure}
\includegraphics[scale=1.6]{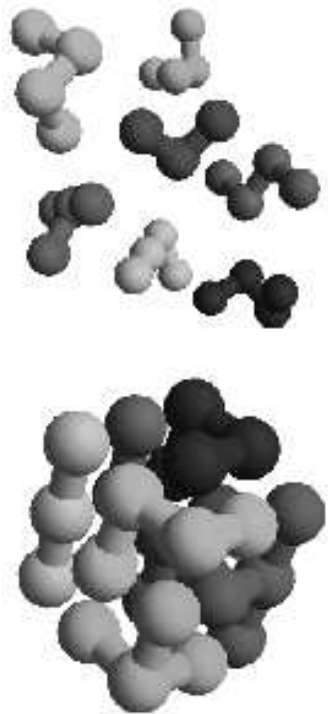}
\caption{\label{fig:19} (Color online) Soma cube components and self-assembled
puzzle solution.}
\end{figure}

The second system is an MD realization of the `Soma cube' \cite{ber82}. This is
an assembly puzzle consisting of seven distinct pieces, each formed from three
or four unit cubes joined along their faces in various fixed configurations;
the pieces can be assembled to form a cube of edge three in 240 different ways.
For the simulation, the puzzle pieces are replaced by rigid bodies made of
spheres located at the cube positions, with attractive forces between spheres
that are adjacent when all pieces fit together in a particular solution of the
puzzle. If the bodies are randomly placed in a large container (with reflecting
walls as before), assigned random initial velocities, their dynamics simulated
using MD, and subjected to a gradually falling temperature, will the puzzle
solution emerge spontaneously? The answer is a qualified yes; in a substantial
proportion of runs the cube assembles itself. Figure~\ref{fig:19} shows the
components and the assembled product, providing yet another example of how, if
interactions are formulated properly, the outcome is correct self-assembly.

\section{Conclusion}

The approach described in this paper involves placing simplified capsomer
elements in a container and following their dynamics; the outcome is a
demonstration that a simple potential energy function, based on structural
considerations, is essentially all that is required to drive the assembly of
the corresponding polyhedral capsid shells. The surprising aspects of the
results -- given the absence of any {\em a priori} theoretical expectation --
are the fast growth rates, high yields, and the avoidance of incorrect
structures. Although the models are not representative of real molecules, if
general principles underlying capsid assembly do exist, simplified systems of
this kind ought to embody their essence.

The advantage of the simplified approach, once its validity and relevance are
confirmed, is that it allows exploration of the salient features of the problem
free of peripheral detail; the influence of shape and interactions can be
examined relatively easily, in contrast to the heavy computational demands of
an explicit atomic representation. While the qualitative benefits are
indisputable, the approach is not intended for quantitative estimation, where
accurate molecular structure and interactions are likely to be important. This
does not preclude basing systematic studies on simple models; it is possible,
for example, to explore the effects of varying relative interaction strengths
to match known binding energies, or to introduce design modifications aimed at
better approximating major structural features of the capsomers. Such projects
await future consideration.

\begin{acknowledgments}
This work was partially supported by research and equipment grants from the
Israel Science Foundation. The author wishes to thank J. Johnson for
introducing him to the subject, and C. Brooks and J. Skolnick for helpful
discussion.
\end{acknowledgments}

\bibliography{passem}

\end{document}